%% file: main.tex
\documentclass[
    twocolumn,
	prd,
    amssymb,
	preprintnumbers,
	secnumarabic,
	nofootinbib,
	superscriptaddress]{revtex4-1}

\pdfoutput=1

\usepackage{graphicx}
\usepackage{enumitem}
\usepackage{latexsym}
\usepackage{amsfonts}
\usepackage{amssymb}
\usepackage{array}
\usepackage{pifont}
\usepackage{color}
\usepackage{xcolor}
\usepackage{amsmath}
\usepackage{slashed}
\usepackage{dcolumn}
\usepackage{verbatim}
\usepackage{float}
\usepackage{multirow}
\usepackage{xspace}
\usepackage[normalem]{ulem}
\usepackage{hyperref}
\usepackage{subfigure}
\usepackage{anyfontsize}
\usepackage{t1enc}
\usepackage{tabularx} 
\usepackage{diagbox}

\definecolor{aquamarine}{rgb}{0.2,0.7,0.6}
\definecolor{cerulean}{RGB}{0,166,214} 
\definecolor{hypershade}{rgb}{0.3,0.3,0.8}
\definecolor{subtlered}{rgb}{0.8,0.3,0.3}

\hypersetup{
     colorlinks   = true,
     citecolor    = Aquamarine,
     urlcolor     = Aquamarine,
     linkcolor    = Aquamarine
}

\usepackage{tikz,xcolor,hyperref}

\definecolor{lime}{HTML}{A6CE39}
\DeclareRobustCommand{\orcidicon}{\hspace{-1mm}
	\begin{tikzpicture}
		\draw[lime, fill=lime] (0,0) 
		circle [radius=0.16] 
		node[white] {{\fontfamily{qag}\selectfont \tiny \,ID}};
		\draw[white, fill=white] (-0.0525,0.095) 
		circle [radius=0.007];
	\end{tikzpicture}
	\hspace{-3mm}
}

\foreach \x in {A, ..., Z}{\expandafter\xdef\csname orcid\x\endcsname{\noexpand\href{https://orcid.org/\csname orcidauthor\x\endcsname}
		{\noexpand\orcidicon}}
}

\hypersetup{
  pdfauthor={Nirmal Raj},
  pdftitle={kwik hawking},
  pdfsubject={kwik hawking},
  colorlinks=true,
  citecolor=aquamarine,
  urlcolor=cerulean,
  linkcolor=cerulean
}

\input{universalnewcommands.tex}

\def\RNS{R_{\rm NS}}
\def\MNS{M_{\rm NS}}

\def\mx{m_\chi}

\def\sigmanx{\sigma_{\rm \chi n}}

\allowdisplaybreaks

\begin{document}

\title{Hawking heating of neutron stars by dark matter}

\author{Akash Kumar Saha\orcidA{}}
\email{akashks@iisc.ac.in}

\affiliation{Centre for High Energy Physics, Indian Institute of Science, C. V. Raman Avenue, Bengaluru 560012, India}

\author{Abhishek Dubey\orcidB{}}
\email{abhishekd1@iisc.ac.in}

\affiliation{Centre for High Energy Physics, Indian Institute of Science, C. V. Raman Avenue, Bengaluru 560012, India}

\author{Nirmal Raj\orcidC{}}
\email{nraj@iisc.ac.in}

\affiliation{Centre for High Energy Physics, Indian Institute of Science, C. V. Raman Avenue, Bengaluru 560012, India}

\date{\today}

\begin{abstract}
Interactions with particle dark matter could brighten old, isolated neutron stars to thermal luminosities detectable at current and next-generation telescopes. 
We present a novel mechanism for such signals.
Non-annihilating (e.g., asymmetric) dark matter capturing in a neutron star could form a small black hole in its core, which could then rapidly evaporate away. 
If black holes form and evaporate within the cooling timescale of the neutron star, periodic episodes of black hole evaporation could impart a steady-state stellar luminosity, providing a source of heat additional to the kinetic energy of dark matter during capture. 
Consequently, we obtain sensitivities to dark matter-nucleon cross sections that are stronger than that from dark kinetic heating by a factor of a few for > $10^4$~GeV (> $10^{10}$~GeV) mass of spin-0 (spin-1/2) dark matter. 
\end{abstract}

\maketitle

\section{Introduction}

\begin{figure}[!h]
	\begin{center}	\includegraphics[width=\columnwidth]{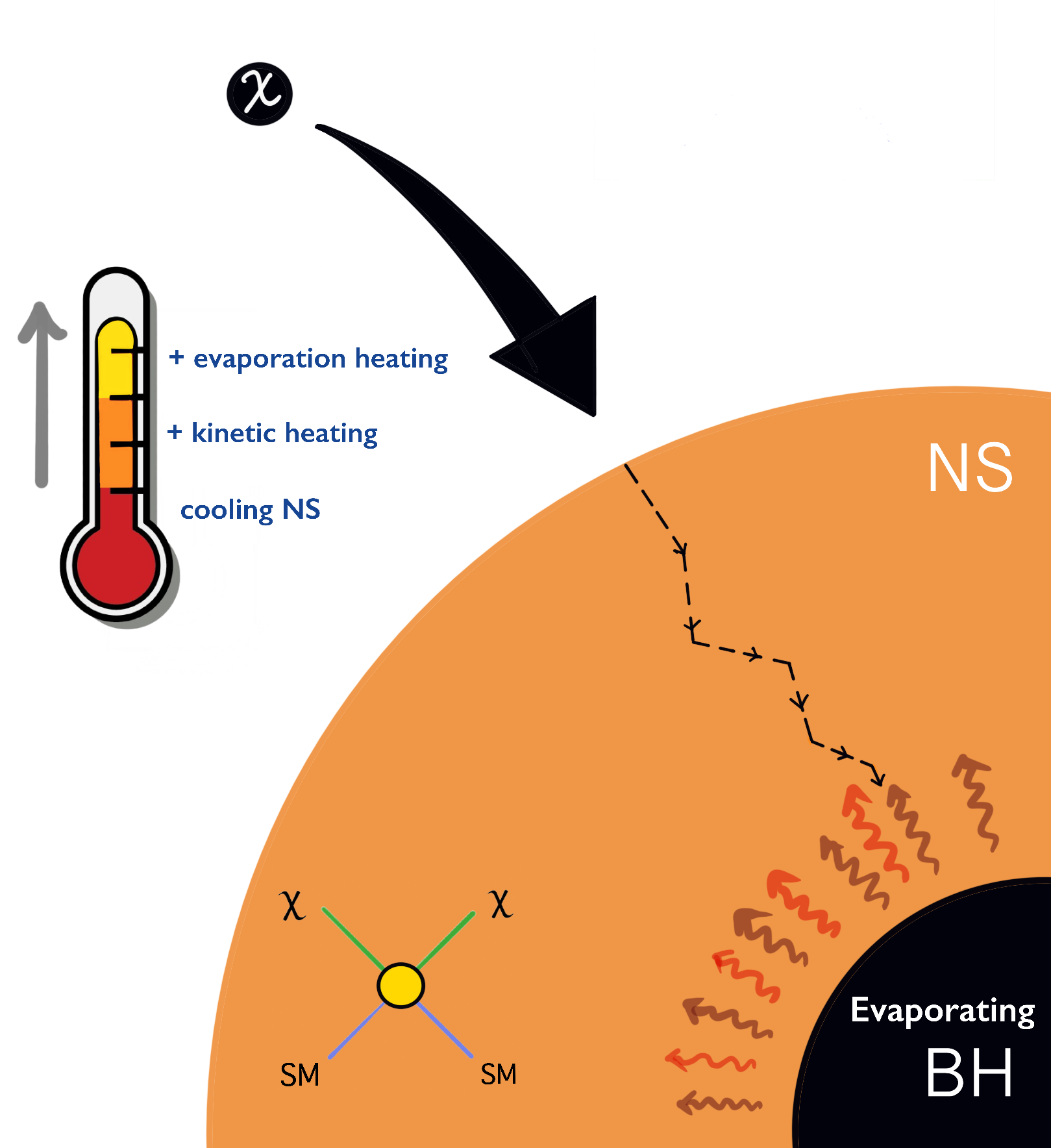}
	\caption{Illustration (not to scale) of our premise. 
     Non-annihilating DM captures in, thermalizes with, and collapses to form a rapidly evaporating BH in the NS. 
     The evaporation products heat the NS core, and the cycle continues. 
     The resulting near-steady-state luminosity of the NS can be measured by infrared telescopes.}
	\label{fig:cartoon} 
     \end{center}
\end{figure}

Unmasking the mysterious dark matter of the cosmos in one of the several experimental efforts around the globe is a keenly awaited event\,\cite{Cirelli:2024ssz, Slatyer:2017sev,Lisanti:2016jxe, Lin:2019uvt}. 
Meanwhile, compact stars offer a most promising setting for testing interactions between states of dark and visible matter~\cite{BramanteRajCompactDark:2023djs,snowmass:ExtremeBaryakhtar:2022hbu}.  
In this {\em Letter}, we will show that particle dark matter (DM) that captures in neutron stars (NSs) and goes on to form black holes (BHs) would steadily deposit its mass energy via rapid Hawking evaporation of the BHs, and overheat the NSs to detectable luminosities; see Fig.~\ref{fig:cartoon}.
This closes an important gap in the literature -- see Table~\ref{tab:litsit} -- and extends the parametric sensitivity of non-annihilating DM across several orders of magnitude in DM mass. 
Our results are summarized in Fig.~\ref{fig:limits}.

\begin{table*}[]
    \begin{tabularx}{0.88\textwidth}{ c c c c} \hline\hline
  \backslashbox{scenario}{phenomenon}    & destroy NS  &  DM kinetic energy $\ra$ heat & DM mass energy $\ra$ heat \\
       \hline
   annihilating DM   &  \cite{LiuPospelovReddy:2025qco}  & \cite{NSvIR:Baryakhtar:DKHNS}  & \cite{Kouvaris:2007ay} \\
   non-annihilating DM  & \cite{Goldman:1989nd} & \cite{NSvIR:Baryakhtar:DKHNS} & {\bf this work} \\
   \hline
        \end{tabularx}
    \label{tab:litsit}
      \caption{Works that (to our knowledge) first studied phenomena in neutron stars by which interactions with particle dark matter may be probed. 
      Our work fills the gap of non-annihilating dark matter converting its mass to heat -- by forming rapidly evaporating black holes.}
\end{table*}
 
Depending on model details, the infalling flux of DM may heat an NS~\cite{Kouvaris:2007ay,deLavallaz:2010wp,Kouvaris:2010vv, NSvIR:Baryakhtar:DKHNS,NSvIR:Raj:DKHNSOps,NSvIR:Bell2019:Leptophilic,NSvIR:Riverside:LeptophilicShort,NSvIR:Riverside:Leptophiliclong,NSvIR:Bell:ImprovedLepton,NSvIR:GaraniHeeck:Muophilic,NSvIR:SelfIntDM,NSvIR:Hamaguchi:RotochemicalvDM2019,NSvIR:GaraniGenoliniHambye,NSvIR:Queiroz:Spectroscopy,NSvIR:Bell:Improved,NSvIR:Bell2020improved,NSvIR:DasguptaGuptaRay:LightMed,NSvIR:zeng2021PNGBDM,NSvIR:Queiroz:BosonDM,NSvIR:HamaguchiEWmultiplet:2022uiq,NsvIR:HamaguchiMug-2:2022wpz,NSvIR:PseudoscaTRIUMF:2022eav,NSvIR:SNeSBursts:Raj:2023azx,NSvIR:Hamaguchi:VortexCreepvDM2023,NSvIR:ReheatedAll:Raj:2024kjq,NSvIR:NearestPulsars:Bramante:2024ikc,NSvIR:Bell2018:Inelastic,NSvIR:InelasticJoglekarYu:2023fjj,NSvIR:tidalfifthforce:Gresham2022,NSvIR:clumps2021,NSvIR:anzuiniBell2021improved,NSvIR:Marfatia:DarkBaryon,snowmass:ExtremeBaryakhtar:2022hbu,snowmass:Carney:2022gse,NSvIR:IISc2022}, 
form a BH inside it that accretes and destroys it~\cite{Goldman:1989nd,Gould:1989gw,Bertone:2007ae,deLavallaz:2010wp,McDermott:2011jp,Kouvaris:2010jy,Kouvaris:2011fi,Kouvaris:2012dz,Bell:2013xk,Guver:2012ba,Bramante:2013hn,Bramante:2013nma,Kouvaris:2013kra,Bramante:2014zca,Garani:2018kkd,Kouvaris:2018wnh,Lin:2020zmm,Dasgupta:2020mqg,Fuller:2014rza,Bramante:2015dfa,Bramante:2017ulk,Takhistov:2020vxs,Garani:2021gvc,Steigerwald:2022pjo,Bhattacharya:2023stq,Liang:2023nvo,Bramante:2024idl,Basumatary:2024uwo,LiuPospelovReddy:2025qco,snowmass:ExtremeBaryakhtar:2022hbu}, or convert it directly to a BH or a quark star~\cite{Bhutani:2025jfo,silkstrangeletsNS:2010xlt,silkdecay:2014dra,DMann:bubblenucleation:Silk2019}. 
In particular, current understanding is that DM that undergoes negligible or no self-annihilations, such as an asymmetric population, could either heat the NS by transferring kinetic energy alone or form a star-consuming BH.
Limits from the latter apply only up to some DM mass $\mx = \mx^{\rm max}$; the BH mass is inversely proportional to positive powers of $\mx$, and BHs that are too light would evaporate away within the observed lifetimes of old NSs.
Here we turn this into a feature.
For $\mx \gsim \mx^{\rm max}$, rapid evaporation may remove the BH, yet within the NS lifetime another fast-evaporating BH forms, and another, and so on, via the steady influx of DM.
The energy deposited from successive periodic bouts of Hawking radiation would then heat the NS core, which we dub ``Hawking heating''.
As the cooling timescales of NSs are generally large, this phenomenon has the effect of maintaining a nearly steady NS temperature.
Thus non-observations of overluminous NSs at this temperature in upcoming infrared telescopes would restrict these DM models.
At NS geometric cross sections most of the incident mass of DM is ultimately converted to Hawking evaporation, and for lower cross sections the heating luminosities are proportionally lower. 
Thus, our limits are similar to those derived from capture followed by kinetic and/or annihilation heating of NSs, albeit operational at DM masses higher than those from NS $\to$ BH transmutation.
Hawking heating could overheat the Earth~\cite{Acevedo:2020gro} and DM-induced-BH evaporation could trigger Type Ia-like supernovae~\cite{Acevedo:2019gre}.
 DM interactions have been studied in the context of numerous celestial targets~\cite{1987ApJ...321..560G, Hooper:2008cf,IceCube:2011aj,Bernal:2012qh,IceCube:2016dgk,Kopp:2009et,IceCube:2021xzo,Super-Kamiokande:2015xms,Gupta:2022lws,Bose:2021cou,Maity:2023rez, Chen:2023fgr, Ellis:2021ztw,Peled:2022byr,Bhattacharya:2024pmp,John:2024thz,Bose:2024wsh,Gupta:2025jte,Krishna:2025ncv,Chu:2024gpe,Leane:2024bvh,Acevedo:2020gro,Croon:2023bmu,Chu:2024gpe,Chauhan:2023zuf,Hong:2024ozz,Beram:2025dly,Berlin:2024lwe,Nguyen:2025ygc,Croon:2025aof,Hooper:2025ohk,IceCube:2025fcu,Acevedo:2025rqu,Gustafson:2025ypo,1987ApJ...321..571G,Chauhan:2016joa,IceCube:2024yaw,Bramante:2022pmn,Renzi:2023pkn,Garani:2019rcb,Leane:2021tjj,French:2022ccb,Blanco:2023qgi,Robles:2024tdh,Bramante:2019fhi,Adler:2008ky,Blanco:2024lqw,Banks:2024eag,Leane:2020wob,Bhattacharjee:2022lts,Acevedo:2024zkg,Leane:2022hkk,Cappiello:2023hza,Linden:2024uph, Ray:2023auh,Ding:2025rlb,Phoroutan-Mehr:2025hjz,Blanco:2025wpo,Cappiello:2025yfe,Bramante:2023djs,Baryakhtar:2022hbu,Baryakhtar:2017dbj,Raj:2017wrv,brayeur2012enhancement,Acevedo:2019agu,Joglekar:2019vzy, Joglekar:2020liw,Garani:2020wge, Busoni:2021zoe,Maity:2021fxw,Bramante:2021dyx, Bose:2021yhz,Nguyen:2022zwb,Bell:2023ysh,Bhattacharjee:2023qfi,Bhattacharjee:2024pis,Graham:2018efk,Acevedo:2019gre,Bell:2021fye,Acevedo:2023xnu,Garani:2023esk,Raj:2023azx,Raj:2024kjq,Acevedo:2024ttq,Basumatary:2024uwo,Bramante:2024ikc,Sen:2024oes,Dutta:2024vzw,Bell:2025acg,Araujo:2025kpn,Davoudiasl:2025gxn}.

\begin{figure*}
    \centering
    \includegraphics[width=0.45\textwidth]{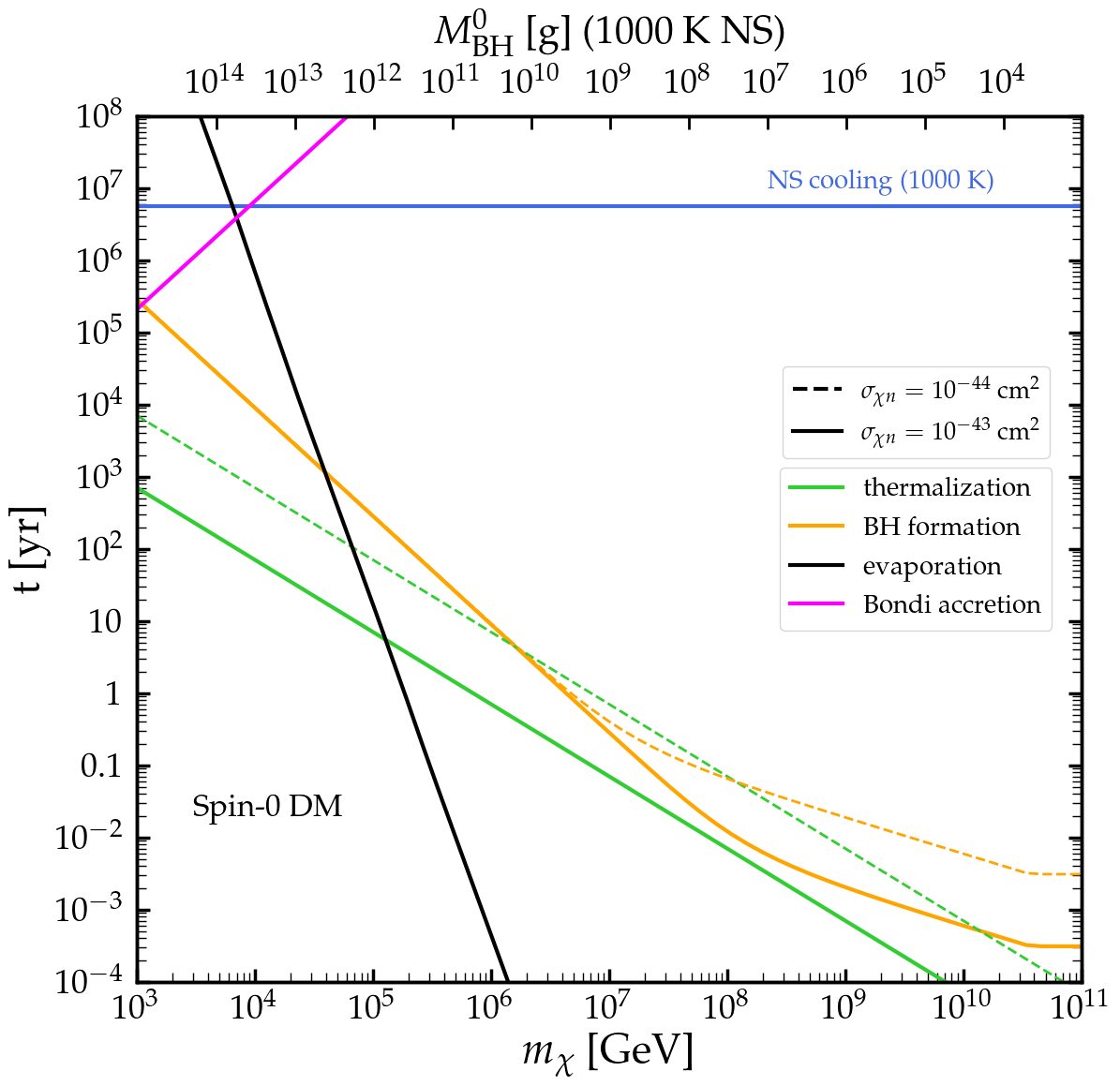}
    \includegraphics[width=0.45\textwidth]{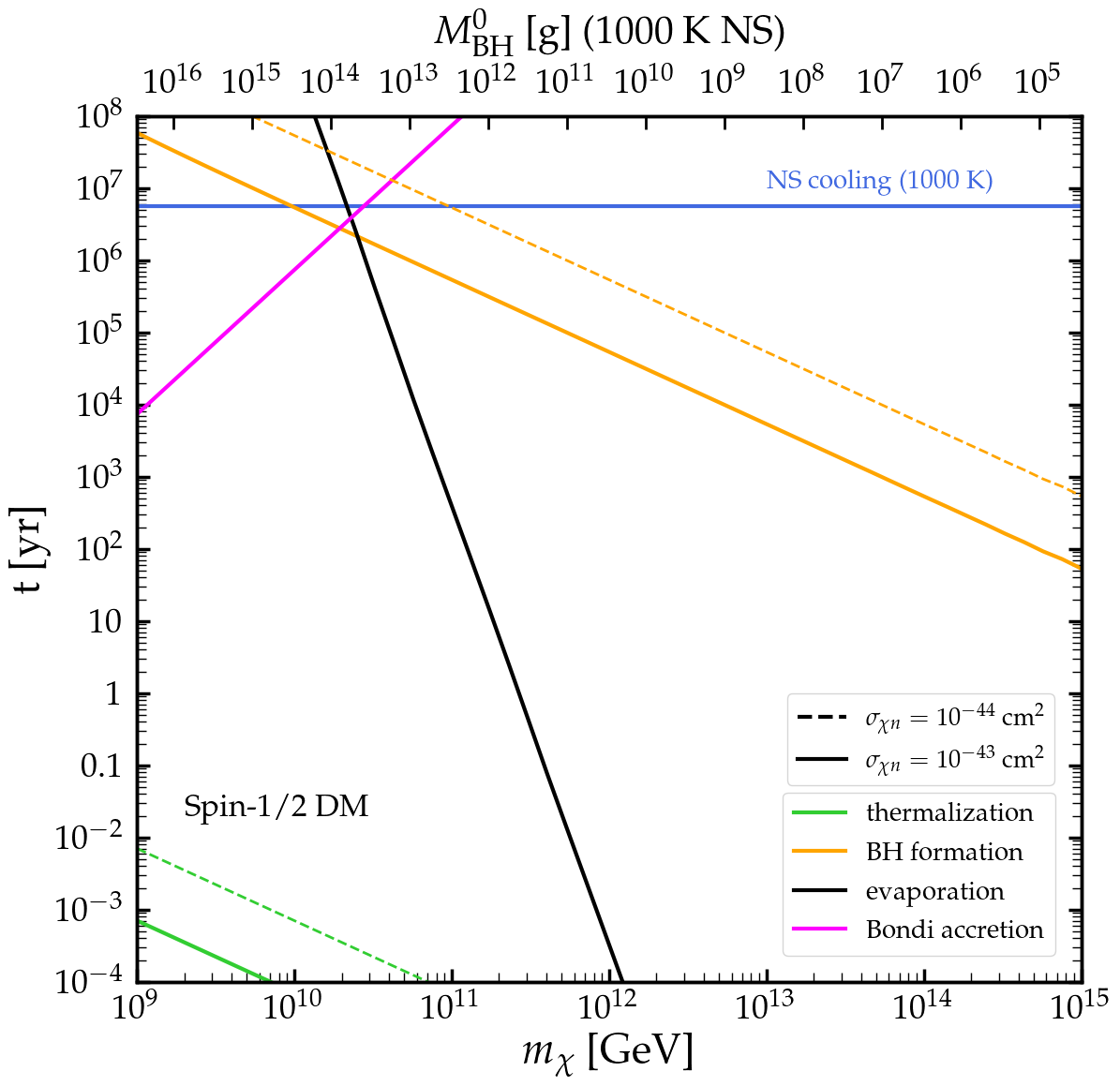}
    \caption{Timescales for various processes in a 1000 K NS as a function of DM mass (and the corresponding $M_{\rm BH}^0$ in the top frame ticks).
    Across the DM masses for which $t_{\rm evap} < t_{\rm Bondi}$, the time for a black hole for form and decay ($t_{\rm th} + t_{\rm BH} + t_{\rm evap}$) is seen to be generally smaller than the NS cooling timescale. 
    In these regions the NS may be observed to glow at a near-constant temperature.
    See Sec.~\ref{subsec:signatures} for further particulars.
   }
    \label{fig:tvmx}
\end{figure*}




\section{Set-up and Signals}
\label{sec:setup}

\subsection{Dark matter capture in neutron star, black hole formation and evolution}
\label{subsec:captureBHform}

We assume that DM (with negligible or no
self-annihilations) is of spin 0 or 1/2 and that it scatters on nucleons via velocity-independent operators. 
The rate of number capture in an NS of mass $\MNS$ and radius $\RNS$ is then given by~\cite{NSMultiscat:Bramante:2017xlb,Ilie:2020vec}

\bea
\label{eq:capture}
&& C_\chi = \sum_{N=1}^\infty C_N~, \\
\nn && C_N = p_N(\tau) \times \frac{\rho_\chi}{\mx} \times \pi (\RNS \gamma)^2 \times \frac{\sqrt{6}}{3 \sqrt{\pi}\, \bar{v}}  \\ 
\nn && \times\bigg[ ( 2\bar{v}^2 + 3 v_{\text{esc}}^2 ) - ( 2\bar{v}^2 + 3 v_N^2)
\exp\!\left( - \frac{3 \left( v_N^2 - v_{\text{esc}}^2 \right)}{2 \bar{v}^2} \right) \bigg],\\
\nn  && p_N(\tau) = 2 \int_0^1 dy \, \frac{y e^{-y\tau} (y \tau)^N}{N!}\,\,,
\eea
where $C_N$ and $p_N$ are the capture rate and probability for $N$ scatters for an optical depth $\tau = 3\sigmanx/2\sigma_{\rm geo}$ and cosine of the DM incident angle $y$, with DM-nucleon scattering cross section $\sigmanx$ and the geometric cross section $\sigma_{\rm geo} = \pi \RNS^2/N_n$ for $N_n$ constituent neutrons;
for NS surface escape speed $v_{\rm esc} = \sqrt{2G\MNS/\RNS}$, we have $\gamma = (1- v_{\rm esc}^2)^{-1/2}$; $v_N=v_{\rm esc}(1-\beta_+/2)^{-N/2}$ with $\beta_+=4m_\chi m_N/(m_\chi+m_N)^2$ for neutron mass $m_N$; $\rho_\chi$ and $\bar{v}$ are the ambient DM density and average speed, taken as 0.4~GeV/cm$^3$ and 220~km/s in the solar vicinity.
In practice, we cut off the summation at some large $N$ beyond which $p_N \to 0$.

Following capture, continued scattering shrinks the orbits of DM particles to within the stellar volume, following which, via further repeated scattering, they thermalize with the NS core of temperature $T_{\rm NS}$ over a timescale~\cite{NSvIR:ThermaliznBertoni:2013bsa,NSvIR:GaraniGuptaRaj:Thermalizn} 
\bea
\nn     t_{\rm th} &=& 7.5 \times 10^7 \,{\rm yr}\,\frac{m_\chi m_N}{(m_\chi+m_N)^2}\\ 
 &&\times  \left(\frac{10^{-45}\,{\rm cm^2}}{\sigmanx}\right)  \left(\frac{10^3 \,{\rm K}}{T_{\rm NS}}\right)^2,
    \label{eq:tthermal}
\eea
settling in a thermal volume of radius
\bea
    r_{\rm th}= 6.5\times10^{-4}\, {\rm cm} \left(\frac{10^9\, {\rm GeV}}{m_\chi}\right)^{1/2}\left(\frac{T_{\rm NS}}{10^3 \,{\rm K}}\right)^{1/2} \\ \nonumber
   \times\left(\frac{1.4\times10^{15 }{\rm g/cm^3}}{\rho_{\rm NS}}\right)^{1/2}\,,
   \label{eq:rth}
\eea
where $\rho_{\rm NS}$ is the NS central density.
Accounting for the potential energy of DM particles from both NS gravitational and self-gravitational potentials,
the DM sphere collapses via a Jeans instability when the net mass of DM exceeds~\cite{Acevedo:2020gro,Robles:2025dlv} 
\bea
 \nn    M_\chi^{\rm self}
     =&& 5.8\times10^{29}\, {\rm GeV}\, \bigg(\frac{10^9 \ {\rm GeV}}{m_\chi}\bigg)^{3/2}\bigg(\frac{T_{\rm NS}}{10^3 \,{\rm K}}\bigg)^{3/2} \\ 
   &\times& \bigg(\frac{1.4\times10^{15} \ {\rm g/cm^3}}{\rho_{\rm NS}}\bigg)^{1/2}\,.
   \label{eq:Mxself}
\eea
Now the Chandrasekhar limit of the DM sphere is given by\,\cite{McDermott:2011jp, Kouvaris:2010jy, Robles:2025dlv}
\bea
M_\chi^{\rm Ch} =   \begin{cases}
     9.5 \times 10^{28}~{\rm GeV} \left(\frac{10^9 \, \rm GeV}{m_{\chi}}\right)\,~,~~{\rm spin}~0\,,\\
    1.6 \times 10^{40}~{\rm GeV} \left(\frac{10^9 \, \rm GeV}{m_{\chi}}\right)^2\,,~{\rm spin}~1/2~.
      \end{cases}
\label{eq:MCh}
\eea
A BH is formed in the NS core once a mass $M^0_{\rm BH} = \max[M_\chi^{\rm self},M_\chi^{\rm Ch}]$ is reached by the accumulation of DM, which takes a time $t_{\rm BH} = M_{\rm BH}^0/(\mx C_\chi)$. 
(For spin-0 DM, a Bose-Einstein condensate (BEC) could form and non-trivially affect the dynamics of BH formation~\cite{Kouvaris:2011fi}. 
We will work under the simplifying assumption that no BEC forms, as is possible if, e.g., the BH evaporates to states that heat the DM condensate~\cite{McDermott:2011jp} or self-interactions are present~\cite{Bramante:2013hn}.) 
The BH then Bondi-accretes NS matter and Hawking-evaporates, with its mass evolution following
\bea
 \nn   \dot{M}_{\rm BH} &=& \dot{M}_{\rm Bondi} - \dot{M}_{\rm evap} \,\\
    &=& \frac{\pi \rho_{\rm NSc} G^2 M_{\rm BH}^2}{c_s^3} - \frac{P(M_{\text{BH}})}{G^2 M_{\rm BH}^2}~,    
    \label{eq:BHevolution}
\eea
with the NS material sound speed $c_s = 0.17c$ and $P(M_{\text{BH}})$ the Page factor accounting for the number of particle species emitted and gray-body corrections~\cite{Page:1976df,MacGibbon:1991tj}, computed here using \texttt{BlackHawk}~\cite{Arbey:2019mbc,Arbey:2021mbl}.
This factor increases from $P (M_{\rm BH}) = 1/1135\pi$ for $M_{\rm BH} > 10^{17}$~g (when only photons and neutrinos are emitted) to  $P (M_{\rm BH}) = 1/74\pi$ for $M_{\rm BH} \ll 10^{10}$~g (when all SM states are emitted).\footnote{The evaporation is suppressed at a sub-10\% level by Pauli-blocking of the fermions emitted~\cite{Autzen:2014tza}, thereby negligibly affecting the minimum $\mx$ we can probe.}
We can now define the accretion and evaporation timescales as $t_{\rm Bondi} = \int_{M^0_{\rm BH}}^{M_{\rm NS}} dM_{\rm BH}/\dot{M}_{\rm Bondi}$
and 
$t_{\rm evap} = - \int_{M^0_{\rm BH}}^0 dM_{\rm BH}/\dot{M}_{\rm evap}$.
These become equal for $M^0_{\rm BH} = [3 c_s^3 P(M_{\rm BH})/(\pi G^4 \rho_{\rm NSc})]^{1/4} = 5.6 \times 10^{13}$~g, corresponding to the timescale of $t_{\rm BondEva} = 3.9 \times 10^6$~yr.
For $M_{\rm BH}^0$ smaller than this value ($\Rightarrow$ higher $\mx$), the BH is smaller than the neutron de Broglie wavelength so that quantum effects inform the BH accretion rate, generally getting smaller than the Bondi rate~\cite{Robles:2025dlv}. 
We will find that this is irrelevant for our analysis and in fact strengthens our requirement of $t_{\rm evap} \ll t_{\rm Bondi}$, thus neglect quantum effects here.

\begin{figure*}
	\begin{center}	
    \includegraphics[width=\columnwidth]{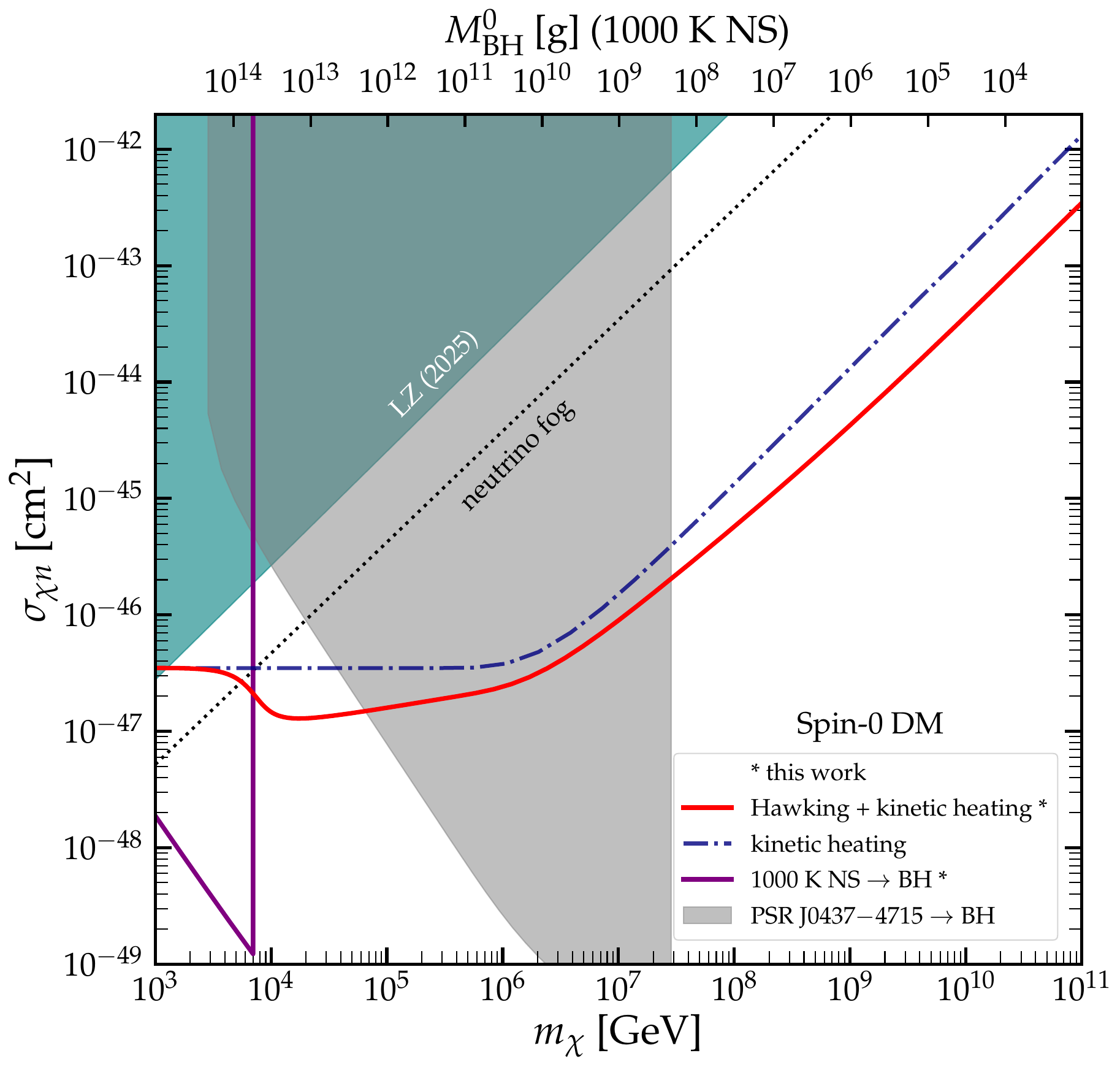}~~\includegraphics[width=\columnwidth]{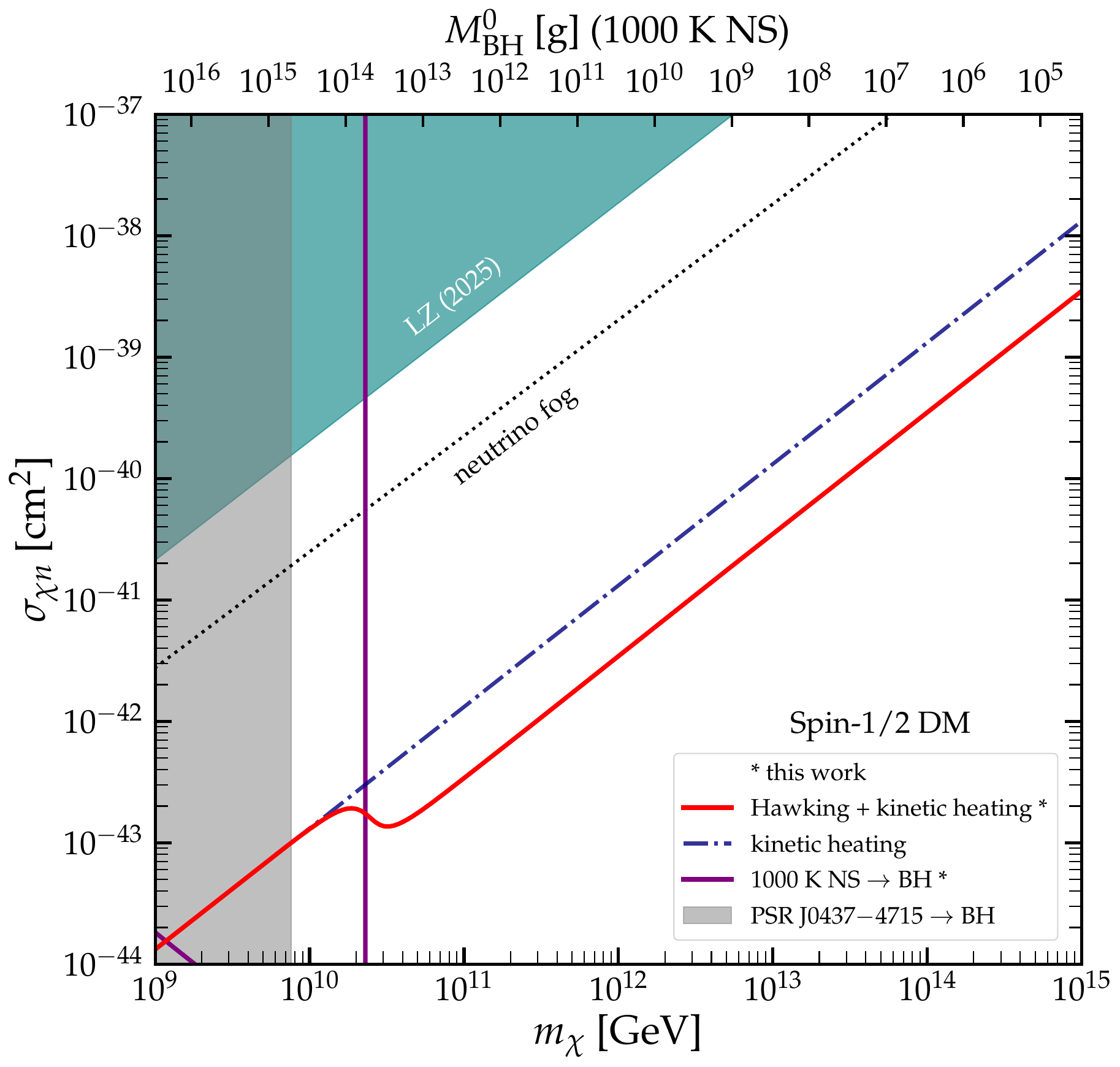}~~\\	
		\caption{ 
        Sensitivities of Hawking heating from the (non-)observation of a 1000~K neutron star to DM-neutron scattering cross section as a function of DM mass (and the corresponding $M_{\rm BH}^0$ in the top frame ticks).
         Also shown for comparison are sensitivities from just kinetic heating; the region that would be probed by observing a 1000~K NS from its not having turned into a black hole and current analogous limits from the existence of the $2 \times 10^6$~K PSR J0437$-$4715 as taken from Ref.~\cite{Bramante:2015dfa,Bramante:2017ulk}; spin-independent direct detection limits from LZ~\cite{LZ:SS:2024zvo}  and sensitivities to the impending neutrino background~\cite{Billard:2013qya,OHare:2021utq}.
      See Sec.~\ref{sec:results} for further details.
        }
		\label{fig:limits}
	\end{center}	
\end{figure*}

\subsection{Observational signatures}
\label{subsec:signatures}

In Fig.~\ref{fig:tvmx} we show the timescales $t_{\rm th}$, $t_{\rm BH}$, $t_{\rm evap}$, $t_{\rm Bondi}$ as a function of $\mx$ for spin-0 and spin-1/2 DM for $\sigmanx = 10^{-44}$~cm$^2$ and $10^{-43} $~cm$^2$, assuming a benchmark NS with $\MNS = 1.5~M_\odot$ and $\RNS = 10~$km.
We have also shown the time for an NS to cool to 1000~K, $t_{\rm NS}^{\rm cool} \simeq 6$~Myr, as obtained from the so-called minimal cooling paradigm~\cite{coolingminimal:Page:2004fy} with analytic expressions for the NS cooling curve from Ref.~\cite{coolinganalytic:Ofengeim:2017cum}.
The time taken for the formation and evaporation of a BH is $t_{\rm tot} = t_{\rm th} + t_{\rm BH} + t_{\rm evap}$.
For spin-0 DM, across the range of $\mx$ corresponding to $t_{\rm evap} < t_{\rm Bondi}$, we see that $t_{\rm tot} \ll t_{\rm NS}^{\rm cool}$. 
This implies that before the energy deposited as heat in the NS by evaporation products is radiated away from the NS surface, the next cycle of BH formation + decay occurs.
Effectively, the NS appears to be in thermal equilibrium over times $> \Oc(t_{\rm cool}^{\rm NS})$.
(We note here that the thermal conduction time across the core is $\Oc$(yr) $\ll t_{\rm NS}^{\rm cool}$~\cite{Cumming:2016weq}.)
For $\mx \gsim 10^5~$GeV, $t_{\rm evap}$ is smaller than the other timescales, and the interval between energy deposition bursts from BHs in the NS  $\Delta t = t_{\rm tot}$.  
For $10^{4}~{\rm GeV} \lsim \mx \lsim 10^{5}~{\rm GeV}$, $t_{\rm evap} > t_{\rm BH}$, implying that multiple BHs may form from the steadily accumulating DM in the time it takes for a single BH to evaporate.
We expect these BHs to co-exist and evolve independently without merging: we do not expect DM collapse to form BHs at the same spot with pinpoint accuracy given the deviations from spherical symmetry of the DM volume from NS spin, capture dynamics, DM-NS thermalization history, and so on.
E.g., for $\mx = 3 \times 10^{4}~$GeV, the Schwarzschild radius of a newly formed BH is $10^{-14}$~cm, whereas the thermal radius of the DM that sourced it is $10^{-1}$~cm (Eq.~\eqref{eq:rth}).
Therefore in this parametric region the interval between energy deposition bursts from BHs is essentially $\Delta t= t_{\rm BH}$. 
For spin-1/2 DM, $t_{\rm tot} < t_{\rm cool}$ for $\sigmanx = 10^{-43}$~cm$^2$, but for smaller $\sigmanx$ we can see that $t_{\rm tot} > t_{\rm cool}$ across a small $\mx$ range.
In this case other future observational signatures may come into play; see Sec.~\ref{sec:discs}.

In practice, we set limits by requiring the DM kinetic and Hawking heating luminosity of the NS to be smaller than its cooling rate:
\bea
\nn (\dot{E}_{\rm kinetic} &+& \dot{E}_{\rm Hawking}) < \dot{E}_{\rm cool}, \\
\dot{E}_{\rm kinetic} &=& (\gamma -1)\,m_\chi C_\chi~, \   \dot{E}_{\rm Hawking} = \frac{M_{\rm BH}^0}{\Delta t}, \nonumber\\
  \dot{E}_{\rm cool} &=& 4\pi\sigma_{\rm SB} \RNS^2 T_{\rm NS}^4~,
\label{eq:limitcriterion}
\eea
where $\gamma=1.34$ for our benchmark NS and $\sigma_{\rm SB}$ is the Stefan-Boltzmann constant. 
Under equilibrium, $\dot{E}_{\rm kinetic}$ imparts a maximum blackbody temperature of 1750~K (corresponding to all the incident DM converting to heat), and $\dot{E}_{\rm kinetic} + \dot{E}_{\rm Hawking}$ to a maximum of 2480~K~\cite{Baryakhtar:2017dbj}.
For these temperatures, photon emission is expected to dominate NS cooling; moreover, the internal and surface temperatures are expected to be equal~\cite{BramanteRajCompactDark:2023djs}.
We will display limits for $T_{\rm NS} = 1000$~K, implying $\dot{E}_{\rm cool} = 4.4 \times 10^{23}$~GeV/s for our benchmark NS; 
an isolated NS of this luminosity can be observed with $10^5$~s of exposure time at the James Webb Space Telescope, Extremely Large Telescope or Thirty Meter Telescope~\cite{NSvIR:ReheatedAll:Raj:2024kjq}.

\section{Results}
\label{sec:results}

In Fig.~\ref{fig:limits} we display our sensitivities in the $\sigmanx$-$\mx$ plane using the criterion in Eq.~\eqref{eq:limitcriterion} and assuming that kinetic + Hawking heating imparts $T_{\rm NS} = 1000$~K; the $\sigmanx$ limit scales as $T_{\rm NS}^{4}$ (from Eqs.~\eqref{eq:capture} and \eqref{eq:limitcriterion}) for $T_{\rm NS} \leq 2480$~K, the maximal value.  
In the top $x$ axis we mark the $\{M^0_{\rm BH}\}$ corresponding to the $\{\mx\}$ in the bottom $x$ axis.
For comparison we also show the reach of purely kinetic heating by DM that would impart $T_{\rm NS} = 1000$~K, and a region that would be excluded by the continued existence of an NS (by avoiding NS$\to$BH transmutation via DM capture) whose temperature would be measured as 1000~K in the future.
One such NS is the coldest known, PSR J2144$-$3933, on whose temperature Hubble Space Telescope observations place an upper limit of about 30,000~K~\cite{coldestNSHST}, while predicted by minimal cooling models to be at $\Oc(100)$~K~\cite{NSvIR:ReheatedAll:Raj:2024kjq}.
This limit is cut off at a value of $\mx = m_\chi^{\rm Bondi}$ beyond which $t_{\rm Bondi} >$ 300 Myr, the spin-down age of PSR J2144$-$3933. 
Our limits effectively begin at $\mx \gsim m_\chi^{\rm Bondi}$, as this is where the BH's evaporation becomes quicker than its accretion of NS matter.
Our Hawking heating sensitivities are generally stronger than that of kinetic heating as the DM mass energy is additionally deposited in the NS via BH evaporation, analogous to its deposition via self-annihilations of DM as studied in the literature. 
Over a small range of $\mx$ below $m_\chi^{\rm Bondi}$ our limits are still stronger than from pure kinetic heating, as contributions from both   kinetic and evaporation heating account for the NS luminosity.
Above $m_\chi^{\rm Bondi}$ our $\sigmanx$ limits are parallel to the kinetic heating limits for most of our $\mx$ range for spin-1/2 DM and for large $\mx$ for spin-0 DM: as seen in Fig.~\ref{fig:tvmx} $\Delta t$ is mostly dominated by $t_{\rm BH}$, in turn resulting in the total heating rate $\propto \mx C_\chi$ in Eq.~\eqref{eq:limitcriterion}, same as the kinetic heating rate.
Deviations from this trend seen in the spin-0 case arise from $\Delta t$ being dominated by other processes such as DM-NS thermalization  and BH evaporation as seen in Fig.~\ref{fig:tvmx}.

We also show limits from (i) the existence of the 6.7~Gyr-old, $2 \times 10^6$~K-hot PSR~J0437$-$4715\,\cite{Bramante:2015dfa,Bramante:2017ulk}, which now spans higher $\mx$: as seen in Eq.~\eqref{eq:Mxself}, to form a BH of a given mass, $\mx \propto T_{\rm NS}$; and (ii) a direct search on spin-independent scattering by LUX-ZEPLIN~\cite{LZ:SS:2024zvo}, depicting also the neutrino fog for xenon targets~\cite{Billard:2013qya,OHare:2021utq} that represents the ultimate reach of direct detectors.
Our sensitivities are seen to complement these.
In particular, while the NS existence bounds always cut off at some DM mass due to $t_{\rm Bondi}$ being too large and $t_{\rm evap}$ being too small, no such limitation applies to us.

\section{Discussion}
\label{sec:discs}

In our study we had made a number of standard assumptions.
Our NS cores were assumed to consist of nucleons, but exotics such as hyperonic or quark cores could be the DM scattering targets~\cite{Robles:2025dlv}.
DM was assumed to be homogeneously distributed in the halo, but it may be clumped, in which case one more timescale enters our analysis: the interval between DM clump-NS encounters~\cite{NSvIR:clumps2021}.
BH evaporation was assumed to be at the Hawking rate, but it may be significantly slowed down by quantum effects such as memory burden~\cite{Basumatary:2024uwo}, altering our parametric sensitivities.
Spherical Bondi accretion onto the BH was assumed, but this could be modified by non-spherical and quantum effects~\cite{East:2019dxt}.
BH formation via a BEC formed by spin-0 DM would be an interesting study.
We had focused on a particular signature of Hawking heating: a steady-state luminosity picked up by observing a single NS.
However, we may be able to probe cross sections smaller than displayed in our results with other signatures. 
The BH formation interval $\Delta t$ could exceed the cooling timescale $t_{\rm cool}$ (as in Fig.~\ref{fig:tvmx} right panel), in which case a fraction $\Delta t/t_{\rm cool}$ of NSs in a given population may be observed to be overheated in a sky survey, such as described in Ref.~\cite{NSvIR:clumps2021}.
We leave these investigations to future work. 

\section*{Acknowledgments}

We thank 
Debajit Bose,
Joe Bramante, 
Anirban Das,
Deep Jyoti Das,
Ranjan Laha, 
Tarak Nath Maity,\\
 Ranjini Mondol,
and
Todd Thompson
for helpful discussion. 
A.K.S.~acknowledges the Ministry of Human Resource Development, Government of India, for financial support via the Prime Minister’s Research Fellowship (PMRF).
N.R.~acknowledges support from the grant ANRF/ECRG/2024/000387/PMS.


\bibliography{refs}

\end{document}

%% file: universalnewcommands.tex
\newcommand{\gsim}{\gtrsim}
\newcommand{\lsim}{\lesssim}
\newcommand{\ra}{\rightarrow}

\def\Oc{\mathcal{O}}

\newcommand{\beq}{\begin{eqnarray}}
\newcommand{\eeq}{\end{eqnarray}}
\newcommand{\bea}{\begin{eqnarray}}
\newcommand{\eea}{\end{eqnarray}}
\newcommand{\nn}{\nonumber}